\documentclass[12pt]{article}
\title{QCD: Confinement, Hadron  Structure  and DIS}
\author{Yu.A.Simonov\thanks{Invited talk at the International Conference dedicated
to the 90-th birthday of the late Professor I.Ya.Pomeranchuk}\\
 State Research
Center\\Institute of Theoretical and Experimental Physics, \\
Moscow, 117218 Russia}
 \date{}
\newcommand{\beq}{\begin{eqnarray}}
 \newcommand{\eeq}{\end{eqnarray}}
\newcommand{\be}{\begin{equation}}
 \newcommand{\ee}{\end{equation}}

\def\ga{\mathrel{\mathpalette\fun >}}
\def\fun#1#2{\lower3.6pt\vbox{\baselineskip0pt\lineskip.9pt
\ialign{$\mathsurround=0pt#1\hfil ##\hfil$\crcr#2\crcr\sim\crcr}}}

\newcommand{{\SD}}{\rm SD}

\newcommand{\veg}{\mbox{\boldmath${\rm g}$}}

\newcommand{\vex}{\mbox{\boldmath${\rm x}$}}

\newcommand{\ver}{\mbox{\boldmath${\rm r}$}}

\newcommand{\veP}{\mbox{\boldmath${\rm P}$}}
\newcommand{\vep}{\mbox{\boldmath${\rm p}$}}

\newcommand{\vek}{\mbox{\boldmath${\rm k}$}}

\newcommand{\verho}{\mbox{\boldmath${\rm \rho}$}}

\newcommand{\lan}{\langle}
\newcommand{\ran}{\rangle}
\begin{document}
\maketitle

\hspace*{5cm} {\it To the   memory of my first teacher

\hspace*{5cm} Isaak Yakovlevich Pomeranchuk.}

%\hspace*{5cm}The Physicist, The Teacher, The Man.}

%\vspace{3mm}

\vspace{10mm}

\begin{abstract}

The main features of QCD, e.g. confinement, chiral symmetry
breaking, Regge trajectories are naturally and economically
explained in the framework of the Field Correlator Method (FCM).
The same method correctly predicts the spectrum of hybrids and
glueballs. When applied to DIS and high-energy scattering it leads
to the important role of higher Fock components in the Fock tower
of moving hadron, containing primarily gluonic excitations.
\end{abstract}

\section{Introduction}

The QCD is the (only) internally selfconsistent theory, defined by
the only scale (parameter the string tension $\sigma$ or
$\Lambda_{QCD}$ can be used for this purpose), which eventually
explains fundamental structure of baryons and other hadrons and by
this some 98\% of all the mass in our world. The features of QCD
are unique in their complexity: confinement, chiral symmetry
breaking, string structure of hadrons  demonstrating
nonperturbative (NP) interactions, on one hand  and on another
hand the applicability of perturbation theory for large $Q$ and
large momentum processes, where NP effects enter as
corrections.The most popular theoretical approaches to these
phenomena look fragmentary, i.e. magnetic monopoles are used for
confinement, instantons for chiral symmetry breaking and pure
perturbative expansions for high energy processes.

The situation has improved with the introduction of the QCD sum
rules \cite{1}, where NP contributions are encoded in the form of
local condensates. In this talk I shall describe a general method
which allows to consider all NP effects on one ground  --  with
the help of nonlocal Field Correlators (FC) \cite{2}. It will be
argued that  the simplest of these correlators -- the  quadratic
one -- is dominant and is responsible for confinement, chiral
symmetry breaking, and the structure of  meson and baryon spectra.

When supplemented with the Background Perturbation Theory (BPT)
the method allows to treat  valence gluons in the confining
background \cite{3}. This yields the spectrum of hybrids and
glueballs in good agreement with lattice  simulations, and the new
perturbation series for high energy processes, without IR
renormalons and Landau ghost poles. As the new and unexpected
element, it will be argued that hybrids play a more fundamental
role in DIS and high-energy scattering -- as the building blocks
of the colliding hadron wave functions.

The talk is organized as follows. In section 2 FC are introduced
and confinement is related to their properties. In section 3 the
Hamiltonian  for valence components is  written down and
properties of meson spectrum are discussed. In section 4 this
Hamiltonian is extended to the systems with valence gluons and
spectrum of hybrids and glueballs is discussed. In section  5 the
Fock towers are introduced for hadrons and the general matrix
Hamiltonians is written down. In section 6 the role of higher
(hybrid) Fock components in DIS  is discussed and the gluon
contribution to  structure functions and proton spin is
emphasized. The last section concludes the general picture of the
QCD dynamics with the discussion of the selfconsistent calculation
of the FC.

\section{ The QCD vacuum structure. Stochastic $vs$ coherent.}

The basic quantity which defines the vacuum structure in QCD is
the field correlator (FC)
\be
D^{(n)}{(x_1,...x_n)} \equiv \lan F_{\mu_1\nu_1} (x_1)
\Phi(x_1,x_2) F_{\mu_2\nu_2}(x_2) \Phi(x_2, x_3) ...
F_{\mu_n\nu_n} (x_n) \Phi(x_n, x_1)\ran,\label{a1}\ee $$
\Phi(x,y)=P\exp ig \int^x_y A_\mu(z) dz_\mu.$$ The set of FC
(\ref{a1}) for $n=2,3,...$  gives a detailed characteristic of
vacuum structure, including field condensates (for coinciding
$x_1=x_2=...x_n$). On general grounds one can distinguish two
opposite situations: 1) stochastic vacuum 2) coherent vacuum. In
the first case FC form a hierarchy with the  dominant lowest term
$D^{(2)} (x_1,x_2) =D^{(2)} (x_1-x_2)$, while higher FC are fast
decreasing with $n$.  We shall call this situation the Gaussian
Stochastic Approximation  (GSA). In the second case all FC are
comparable, and expansion of physical amplitudes as the series of
FC is impractical. This is the case for the gas/liquid of
classical solutions, e.g. of instantons, magnetic monopoles etc.
The physical picture behind the situation of nonconverging FC
series is that of  the  coherent lump(s), when all points in the
lump are strongly correlated.

To understand where belongs the QCD vacuum one can start with the
Wilson loop in the representation $D$ of the color group  SU(3),
\be
W_D(C) = \lan tr_D\exp (ig \int_C  dz_\mu  A^a_\mu
T_a^{(D)})\ran\label{a2} \ee

The Stokes theorem and the cluster expansion identity allow  to
obtain the basic equation, which is used  in most applications of
the FCM (for more details see \cite{2})
\be
W(C)= \exp \sum_n \frac{(ig)^n}{n!} \int \tilde
D^{(n)}(x_1,...x_n) d\sigma_{\mu_1\nu_1}(x_1)...
d\sigma_{\mu_n\nu_n} (x_n).\label{a3}\ee

Here integration is performed over the minimal surface $S_{min}$
inside the contour $C$ defined in (\ref{a2}) and $\tilde D_n$ is
the so-called cumulant or the connected correlator, obtained from
the FC Eq.(\ref{a1}) by subtracting all disconnected averages.
From (\ref{a3}) one easily obtains that the Wilson loop has the
area-law asymptotics, $W(C)\sim \exp (-\sigma S_{min})$, for  any
finite number of terms $n_{max}; n\leq n_{max}$ in  the exponent
(\ref{a3}).

The string tension is expressed through $\tilde D^{(n)}$. \be
 \sigma
=\frac12 \int D^{(2)} (x_1-x_2) d^2 (x_1-x_2) +0(\tilde
D^{(n)},n\geq 4 ) = \sigma_2+\sigma_4 +...\label{a4}\ee

Eq.(\ref{a4}) has several  consequences: 1) confinement appears
naturally for $n=2$, i.e. in the GSA 2) the lack of confinement
can be due to vanishing of all FC, or  due to the special
cancellation between the cumulants, as it happens for the
instanton vacuum \cite{4a}, 3) for static quarks in the
representation $D$ of  the color   group SU(3), the string tension
$\sigma^{(D)}_2$ is proportional to the quadratic Casimir factor (
the  Casimir scaling)
\be
\sigma_2^{(D)} = \frac{C_D^{(2)}}{C_D^{(fund)}}
 \sigma_2^{(fund)}, C_D^{(2)} =\frac13 (\mu^2+\mu\nu+\nu^2+3\mu+3\nu).\label{a5}\ee
 However for larger $n,n\geq 4$ the Casimir scaling is violated:
 \be \sigma_n^{(D)}= a_1C_D^{(2)} +a_2 (C^{(2)}_D)^2
 +a_3C^{(3)}_D+...\label{a6}\ee
 It is remarkable that perturbative interaction of static quarks
 $V^{(D)}(r)$ satisfies the Casimir scaling to the order $O(g^6)$
 considered so far \cite{5a},\footnote{The diagrams of the order $O(g^8)$
 violating the Casimir scaling have been found by W.Wetzel (to be published)}
  so the total potential $V^{(D)}(r) =
 V^{(D)}_{pert} (r) + \sigma^{(D)} r+const$ is also Casimir
 scaling, if  GSA works well.

 This picture was tested recently on the lattice \cite{11a} and
 confirmed the Casimir scaling with the accuracy around 1\% in the
 range $0.1\leq r\leq 1.1. $ fm. The full theoretical understanding
 of this fundamental fact is still lacking, both for the
 perturbative
 part and for the string tension. On the  pedestrian level the
 Casimir scaling and the quadratic (Gaussian) correlator
 dominance implies that  the vacuum  is highly stochastic and any
 quasiclassical objects, like instantons, are strongly suppressed
 in the real QCD vacuum. The vacuum consists of small white
 dipoles of the size $T_g$ made of neighboring field strength
 operators.  The smallness of $T_g$ might be an explanation for
 the  Gaussian dominance since higher correlator terms in $\sigma$
 are proportional to $(FT_g^2)^n (T_g^2)^{-1}$, where $F$ is the
 estimate of the average nonperturbative vacuum field, $F\sim 500
 $ (MeV)$^2$.

 Lattice calculations of FC have been done repeatedly during last
 decades, using cooling technic \cite{12a} and with less
 accuracy without cooling \cite{13a}. (Recently another  approach
 based on the  so-called gluelump states  was exploited on the
 lattice \cite{14a} and analytically \cite{15a}, which has a direct
 connection to FC).

The basic result of \cite{12a} is  that FC  consists of
perturbative part $O(1/x^4)$ at  small distances and
nonperturbative part $O(\exp (-x/T_g))$ at larger distances  with
$T_g$ in the range $T_g = 0.2$ fm (quenched vacuum) and $T_g=0.3$
fm ( 2 flavours).

Calculations in \cite{13a} and \cite{14a,15a}, as well as sum rule
estimates \cite{16a} yield a smaller value, $T_g\approx 0.13$ fm
to 0.17 fm. This enables us in what follows to take the limit
$T_g\to 0$ keeping $\sigma = const \approx 0.18 $ GeV$^2$, and
consider $\sigma T^2_g$ as a small parameter of expansion, $\sigma
T^2_g \ll1$. For example, the contribution of higher correlators
in (\ref{a4}) is proportional to $\sigma(\sigma
T^2_g)^{\frac{n}{2}-1}, n=4,6,8,...$

\section{Hamiltonian for valence components}

There are two possible approaches to incorporating nonperturbative
field correlators  in the quark-antiquark (or $3q$) dynamics. The
first has to deal with the effective nonlocal quark Lagrangian
containing field correlators \cite{12b}. From this one obtains
first-order Dirac-type  integro-differential equations for
heavy-light  mesons \cite{12,13}, light mesons  and baryons
\cite{14}. These equations contain the effect of chiral symmetry
breaking \cite{12,12b} which is directly connected to confinement.

The second approach is based on the effective Hamiltonian for any
gauge-invariant quark-gluon system. In the limit $T_g\to 0$ this
Hamiltonian is simple and local, and in most cases when spin
interaction can be considered as a perturbation one obtains
results for the spectra in an analytic form, which is transparent.

For this reason we choose below the second, Hamiltonian approach
\cite{15,16}. We start with the exact Fock-Feynman-Schwinger
Representation for the $q\bar q$ Green's function (for a review
see \cite{17}), taking for simplicity nonzero flavor case $$
 G^{(x,y)}_{q\bar q} =\int^\infty_0 ds_1 \int^\infty_0 ds_2
( Dz)_{xy}(D\bar z)_{xy}e^{-K_1-K_2}$$\be\lan  tr \Gamma_{in}
(m_1-\hat D_1) W_\sigma (C) \Gamma_{out} (m_2-\hat
D_2)\ran_A\label{1}\ee
 where $K_i=\int^{s_1}_0 d\tau_i (m_i+\frac14 (\dot
 z_\mu^{(i)})^2), $ $\Gamma_{in,out}=1, \gamma_5,...$ are meson
 vertices, and $W_\sigma(C)$ is the Wilson loop with spin
 insertions, taken along the contour $C$ formed by paths
 $(Dz)_{xy}$ and $(D\bar z)_{xy}$,
 $$
 W_\sigma (C) =P_F P_A\exp (ig \int_CA_\mu dz_\mu)\times$$
 \be
\times \exp (g\int^{s_1}_0
 \sigma^{(1)}_{\mu\nu} F_{\mu\nu} d\tau_1-g\int^{s_2}_0
 \sigma_{\mu\nu}^{(2)}F_{\mu\nu} d\tau_2).\label{2}\ee
 The last factor in (\ref{2}) defines the spin interaction of
 quark and antiquark. The average $\lan W_\sigma\ran_A$ in
 (\ref{1}) can be computed exactly through field correlators $\lan
 F(1)...F(n)\ran_A$, and keeping only the lowest  one,$\lan F(1)
 F(2)\ran$, which yields according to lattice calculation
 \cite{11a} accuracy around 1\% \cite{5a},
 one obtains
$$ \lan  W_\sigma (C)\ran_A \simeq \exp (-\frac12
[\int_{S_{min}}ds_{\mu\nu}(1)
\int_{S_{min}}ds_{\lambda\sigma}(2)+$$ \be+ \sum^2_{i,j=1}
\int^{s_i}_0
 \sigma^{(i)}_{\mu\nu} d\tau_i\int^{s_j}_0
 \sigma_{\lambda\sigma}^{(j)} d\tau_j]\lan F_{\mu\nu}(1) F_{\lambda\sigma}(2)\ran ).
 \label{3}\ee

 The Gaussian correlator $\lan F_{\mu\nu} (1) F_{\lambda\sigma}(2)
 \ran \equiv D_{\mu\nu,\lambda\sigma} (1,2)$ can be rewritten
 identically in terms of two scalar functions $D(x)$ and $D_1(x)$
 \cite{2}, which have been computed on the lattice \cite{12a} to
 have the exponential form $D(x), D_1(x) \sim \exp (-|x|/T_g$ with
 the gluon correlation length $T_g\approx 0.2 $ fm.

As the next step one introduces the einbein variables $\mu_i$ and
$\nu$; the first one to transform the proper times $s_i, \tau_i$
into the actual (Euclidean) times $t_i\equiv z^{(i)}_4$. One has
\cite{16}
\be
2\mu_i(t_i) =\frac{dt_i}{d\tau_i},~~ \int^\infty_0
ds_i(D^4z^{(i)})_{xy} =const \int
D\mu_i(t_i)(D^3z^{(i)})_{xy}.\label{5}\ee The variable $\nu$
enters in the Gaussian representation of the Nambu-Goto form for
$S_{min}$ and its stationary value $\nu_0$ has the physical
meaning of the energy density along the string. In case of several
strings, as in the baryon case or the hybrid case, each piece of
string has its own parameter $\nu^{(i)}.$

To get rid of the path integration in (\ref{1}) one can go over to
the effective Hamiltonian using the identity
\be
G_{q\bar q} (x,y) =\lan x| \exp (-HT) |y\ran\label{6}\ee where $T$
is the evolution parameter corresponding to the hypersurface
chosen for the Hamiltonian: it is the hyperplane $z_4=const$ in
the c.m. case \cite{16}.

The final form of the c.m. Hamiltonian (apart from the spin and
perturbative terms to be discussed later) for  the $q\bar q$ case
is \cite{16,18} $$ H_0=\sum^2_{i=1} \left(
\frac{m^2_i+\vep^2_i}{2\mu_i} +\frac{\mu_i}{2}\right) + \frac{\hat
L^2/r^2}{2[\mu_1(1-\zeta)^2+\mu_2\zeta^2 + \int^1_0 d \beta (\beta
-\zeta)^2\nu (\beta)]}+$$
\be
+\frac{\sigma^2 r^2}{2} \int^1_0 \frac{d\beta}{\nu(\beta)} +
\int^1_0\frac{\nu(\beta)}{2}d\beta.\label{7}\ee

Here $\zeta= (\mu_1+\int_0\beta \nu d\beta)/ (\mu_1+\mu_2+\int^1_0
\beta \nu d \beta)$ and $\mu_i$ and $\nu(\beta)$ are to be found
from the stationary point of the Hamiltonian
\be
\frac{\partial H_0}{\partial\mu_i}|_{\mu_i=\mu_i^{(0)}} =0,~~
\frac{\partial H_0}{\partial\nu}|_{\nu=\nu^{(0)}} =0.\label{8}\ee

Note that $H_0$ contains as input only $m_1, m_2$ and $\sigma$,
where $m_i$ are current masses defined at the scale  of 1 GeV. The
further analysis is simplified by the observation that for $L=0$
one finds $\nu^{(0)}=\sigma r$ from (\ref{8}) and
$\mu_i=\sqrt{m^2+\vep^2}$, hence $H_0$ becomes the usual
Relativistic Quark Model (RQM) Hamiltonian
\be
H_0(L=0)=\sum_{i=1}\sqrt{m_i^2+\vep^2} +\sigma r. \label{9}\ee

 But $H_0$ is not the whole story,
one should take into account 3 additional terms: spin terms in
(\ref{3}) which produce two types of contributions: self-energy
correction \cite{19} \be H_{self}=\sum^2_{i=1} \frac{\Delta
m^2_q(i)}{2\mu_i},~~ \Delta m^2_q=-\frac{4\sigma}{\pi} \eta
(m_i),~~ \eta(0)=\frac{3}{4} (1+\frac{D_1(0)}{D(0)}) \cong 1 \div
0.9\label{14}\ee and spin-dependent interaction between quark and
antiquark $H_{spin}$ \cite{22} which is entirely described by the
field correlators $D(x), D_1(x) $, including also the one-gluon
exchange part present in $D_1(x)$.

Finally one should take into account gluon exchange contributions,
which can be divided into the Coulomb part $H_{Coul}
=-\frac43\frac{\alpha_s(r)}{r},$ and $H_{rad}$ including
space-like gluon exchanges and perturbative self-energy
corrections (we shall systematically omit these corrections since
they are small for light quarks to be discussed below). In
addition there are gluon contributions which are nondiagonal in
number of gluons $n_g$  and quarks (till now only the sector
$n_g=0$ was considered) and therefore mixing meson states with
hybrids and glueballs \cite{20}; we call these terms $H_{mix}$ and
refer the reader to \cite{20} and the cited there references for
more discussion. Assembling all terms together one has the
following total Hamiltonian in the limit of large $N_c$ and small
$T_g$ (for more discussion see \cite{1b}):
\be
H=H_0+H_{self} +H_{spin} +H_{Coul}+H_{rad} +H_{mix}.\label{15}\ee

We start with $H_0=H_R+H_{string}$. The eigenvalues $M_0$ of $H_R$
can be given with 1\% accuracy by \cite{21}
\be
M^2_0\approx 8\sigma L+4\pi\sigma (n+\frac34)\label{16}\ee where
$n$ is the radial quantum number, $n=0,1,2,...$ Remarkably
$M_0\approx 4\mu_0$, and  for $L=n=0$ one has $\mu_0(0,0)=0.35$
GeV for $\sigma =0.18 $ GeV$^2$, and $\mu_0$ is fast increasing
with growing $n$ and $L$. This fact  partly explains that spin
interactions become unimportant beyond $L=0,1,2$ since they are
proportional to $d\tau_1 d\tau_2\sim \frac{1}{4\mu_1\mu_2}
dt_1dt_2$ (see (\ref{3}) and \cite{1b}). Thus constituent mass
(which is actually "constituent energy") $\mu_0$ is  "running".
The   validity of $\mu_0$ as a socially accepted "constituent
mass" is confirmed by its numerical value given above, the spin
splittings of light and heavy mesons  \cite{23} and by baryon
magnetic moments expressed directly through $\mu_0$, and being in
agreement with experimental values \cite{24}.

\section{Hamiltonian and bound states of valence gluons}

   We now come to the gluon-containing systems, hybrids and
   glueballs. Referring the reader to the original papers
   \cite{9}-\cite{11} one can recapitulate the main results for
   the spectrum. In both cases the total Hamiltonian has the same
   form as in (\ref{15}), however the contribution of corrections
   differs.

   For glueballs it was argued in \cite{11} that $H_0$ (\ref{7})
   has the same form, but with $m_i=0$ and $\sigma\to
   \sigma_{adj}=\frac{9}{4} \sigma$ while $H_{self} =0$ due to
   gauge invariance.

Thus one can retain in (\ref{15})  only two main terms:
$H=H_0+H_{spin}$ while $H_{Coul}$ was argued to be strongly
decreased by loop corrections. The calculation in \cite{11} was
done for two-gluon and three-gluon glueball states and results are
in surprisingly good agreement with lattice data for both systems
(no fitting parameters have been used in \cite{11}).

 We now coming to the next topic of this talk: hybrids
and their role in hadron dynamics. We start with the hybrid
Hamiltonian and spectrum. This topic in the framework of FCM was
considered in \cite{9,10} The Hamiltonian $H_0$ for hybrid looks
like \cite{1b,9,10}
\be
H_0^{(hyb)} =\frac{m^2_1}{2\mu_1} + \frac{m^2_2}{2\mu_2} +
\frac{\mu_1+\mu_2+\mu_g}{2} + \frac{\vep^2_\xi+\vep^2_\eta}{2\mu}
+\sigma\sum^2_{i=1} |\ver_g-\ver_i| + H_{str}.\label{22} \ee

Here $\vep_\xi, \vep_\eta$ are Jacobi momenta of the 3-body
system, $H_{self}$ is the same as for meson, while $H_{spin}$ and
$H_{Coul}$ have different structure \cite{10}.

The main feature of the present approach based on the BPTh, is
that valence gluon in the hybrid is situated at some arbitrary
point on the string connecting quark and antiquark, and the gluon
creates a kink on the string so that two pieces of the string move
independently (however connected at the point of gluon). This
differs strongly from the flux-tube model  where hybrid is
associated with the string excitation  as a whole, but has a
strong similarity to the treatment of gluons in the framework of
the Lund model \cite{30}.

 Results for light and heavy exotic
$1^{-+}$ hybrids are also given in \cite{1b} and are in agreement
with lattice calculations. Typically an additional gluon in the
exotic $(L=1)$ state "weights" 1.2$\div$1.5 GeV for light to heavy
quarks, while nonexotic gluon $(L=0)$ brings about 1 GeV to the
mass of the total $q\bar q g$ system. Let us now consider the
hybrid spectrum in more detail.  First of all we use for that
3-body problem the hyperspherical method, which works with
accuracy of few percent \cite{31,32}. Then the whole spectrum is
classified by the grand angular momentum $K=0,1,2,..., $ which is
actually an arithmetic sum of all partial pair angular momenta in
the system $q\bar q g$. The lowest $K=0$ states can be formed from
the $s$-wave $q\bar q$ pair and $s$-wave valence gluon $\veg$,
which gives the $\verho+\veg$ and $\pi +\veg $ systems, and
vectors  imply spin-one particle. In this way one obtains the
classification $$K=0, (\pi+\veg) -1^{+-}, (\verho+\veg) - (2^{++},
1^{++}, 0^{++})$$
 $$K=1, (\pi+(\nabla\times\veg)) -1^{--}, (\verho+(\nabla\times \veg)) -
(2^{-+}, 1^{-+}, 0^{-+}).$$

The eigenvalues of the Hamiltonian (\ref{22}) are easily obtained
for the light quarks using the hypercentral (lowest $K$)
approximation -- $$M(K=0)=1.872 {\rm ~GeV}$$ $$M(K=1)=2.45 {\rm
~GeV}$$ $$M(K=2)=2.90 {\rm ~GeV}$$ $$M(K=3)=3.27 {\rm ~GeV}.$$

Here the (negative) self-energy part of quarks $H_{self} $ is
already added to masses. One has also in addition the color
Coulomb part $H_{Coul}$ and spin-dependent part $H_{spin}$, which
contribute approximately $\lan H_{Coul}\ran \sim -0.2$ GeV and for
spin-spin interaction approximately 0.08 GeV $\left
(\begin{array}{l} -2\\-1\\+1 \end{array}\right)$, where numbers
inside brackets refer to $\vec 1+\vec 1 = \vec 0, \vec 1,\vec 2$
respectively from top to bottom.

As a result, neglecting
 rather small contribution from $H_{rad}$ and $H_{mix}$ in
 (\ref{15}), and the string correction, taking into account the
 moment of inertia of the string \cite{16}, yielding around (-100
 MeV), one has the approximate mass estimates for the lowest mass
 states of $K$ multiplets:
 $$M_{low} (K=0)\cong 1.42 {\rm ~GeV}$$
$$M_{low} (K=1)\cong 1.9 {\rm ~GeV}$$ $$M_{low} (K=2)\cong 2.45
{\rm ~GeV}.$$

The mass value $M_{low} (K=1)$ agrees well with the lattice
results for the mass of the $1^{-+}$ state \cite{33}. The recent
unquenched calculation \cite{34} yields the value which is
somewhat lower.

One should stress that the hybrid states, which start at the mass
around 1.4 GeV, have a high multiplicity which grows exponentially
with mass, as well as excited string states in bossonic string
theory \cite{35}. This fact has  a very important  consequence for
high-energy  processes, where the hybrid excitation is argued to
be the dominant physical mechanism.

 \section{Hamiltonian and Fock states}

 As was mentioned above the QCD Hamiltonian is introduced in
 correspondence with the chosen hypersurface, which defines internal
 coordinates $\{\xi_k\}$ lying inside the hypersurface,
and the evolution parameter, perpendicular to it. Two extreme
choices are frequently used, 1) the c.m. coordinate system with
the hypersurface $x_4=const.$, which implies that all hadron
constituents have the same (Euclidean) time coordinates
$x_4^{(i)}=const, i=1,...n$, 2) the light-cone coordinate system,
where the role of $x_4$ and $x_4^{(i)}$ is played by the $x_+,
x_+^{(i)}$ components, $x_+=\frac{x_0+x_3}{\sqrt{2}}$.

To describe the structure of the Hamiltonian in general terms
 we first assume that the bound valence states exist for
mesons, glueballs and baryons consisting of minimal number of
constituents. To form the Fock tower of states starting with the
given valence state, one can add gluons and $q\bar q$ pairs
keeping the $J^{PC}$ assignment  intact. At this point we make the
basic simplifying approximation assuming that the number of colors
$N_c$ is tending  to infinity, so that one can do for any physical
quantity an expansion in powers of $1/N_c$. Recent lattice data
confirm a good convergence of this expansion for $N_c=3,4,6$ and
all quantities considered \cite{36} (glueball mass, critical
temperature, topological susceptibility etc.).

Then the construction of the Fock tower is greatly  simplified
since any additional $q\bar q$ pair enters with the coefficient
$1/N_c$ and any additional white (e.g. glueball) component brings
in  the coefficient $1/N_c^2$. In view of this in the leading
order of $1/N_c$ the Fock tower is formed by only creating
additional gluons in the system, i.e. by the hybrid excitation of
the original (valence) system. Thus all Fock tower consists of the
valence component and its hybrid equivalents and each line of this
tower is characterized by the number $n$ of added gluons. Then,
the internal coordinates $\{\xi\}_n$ describe coordinates and
polarizations of $n$ gluons in addition to those of valence
constituents.

We turn now to the Hamiltonian $H$, assuming it to be either the
total QCD Hamiltonian $H_{QCD}$, or the effective Hamiltonian
$H^{(eff)}$, obtained from $H_{QCD}$ by integrating out
short-range degrees of freedom. We shall denote the diagonal
elements of $H$, describing the dynamics of the $n$-th hybrid
excitation of $s$-th valence state ($s=m\{f\bar f\}, gg, 3g,
b\{f_1f_2f_3\}$ for mesons, 2-gluon and 3-gluon glueballs and
baryons respectively with $f_i$ denoting flavour of quarks) as
$H^{(s)}_{nn}$. For nondiagonal elements we confine ourselves to
the lowest order operators $H^{(s)}_{n,n+1}$ and $H^{(s)}_{n-1,
n}$ describing creation or annihilation of one additional gluon,
viz.
\be
H_{\bar q qg}=g\int\bar q \bar (\vex, 0)\hat a (\vex,0 )) q(\vex,
0) d^3x\label{10a}\ee
\be
H_{g2g} =\frac{g}{2} f^{abc}\int (\partial_\mu
a^a_\nu-\partial_\nu a^a\mu) a^b_\mu a^c_\nu d^3 x,\label{11a}\ee
 and we disregard for simplicity the terms $H_{g3g}$.

 As it is clear from (\ref{10a}), (\ref{11a}), the first operator
refers to the gluon creation from the   quark line, while the
second refers to the creation of 2 gluons from the gluon line. In
what follows we shall be mostly interested in the first operator,
which yields dominant contribution at large  energies, and
physically describes addition of   one last cross-piece to the
ladder of gluon exchanges between quark lines, while (\ref{11a})
corresponds in the same ladder to the $\alpha_s$ renormalization
graphs, and to the graphs with creation of additional gluon line.

The effective Hamiltonian in the one-hadron sector can be written
as follows
\be
\hat H=\hat H^{(0)} +\hat V\label{12a}\ee where $H^{(0)}$ is the
diagonal matrix of operators,
\be
H^{(0)}= \{ H_{00}^{(s)}, H_{11}^{(s)}, H_{22}^{(s)}, ...
\}\label{13a}\ee
 while $\hat V$ is the sum of operators (\ref{10a}) and (\ref{11a}),
 creating and annihilating one gluon. In (\ref{13a}) $H^{(s)}_{nn}$
 is the Hamiltonian operator for what we call the "n-hybrid", i.e.
 a bound state of the system, consisting of $n$ gluons together
 with the particles of  the valence  component. In this way the
 $n$-hybrid for the valence $\rho$-meson is the system consisting
 of $q\bar q$ plus $n$ gluons "sitting" on the string connecting
 $q$ and $\bar q$.

 Before applying the stationary perturbation theory in $\hat V$ to
 the Hamiltonian (\ref{12a}), one should have in mind that there
 are two types of excitations  of the ground state valence Fock
 component: 1) Each of the operators $H_{nn}^{(s)}, n=0,1,...$ has
 infinite amount of excited states, when radial or orbital motion
 of any degree  of freedom is excited, 2) in addition one can add
 a gluon, which means exciting the string and this excitations due
 to the operator $\hat V$ transforms the $n-th$ Fock component
 $\psi_n^{(s)}$
 into $\psi^{(s)}_{n+1}$.

 The wave equation for the Fock tower $\Psi_N\{P,\xi\}$ has the
 standard form
 \be
 \hat H\Psi_N=(\hat H^{(0)}+\hat V) \Psi_N=E_N\Psi_N,\label{14a}\ee
 or in the integral form
 \be
 \Psi_N= \Psi_N^{(0)} - G^{(0)} \hat V \Psi_N \label{15a}\ee
 where $G^{(0)}$ is diagonal in Fock components,
 \be
 G^{(0)}(E) =\frac{1}{\hat H^{(0)}-E},~~
 G^{(0)}_{nm}(E) =\delta_{nm} \frac{1}{ H^{(s)}_{nn}
 -E},\label{16a}\ee
 and $\Psi_N^{(0)}$ is the eigenfunction of $\hat H^{(0)}$,
 \be
 \hat H^{(0)} \Psi_N^{(0)}= E^{(0)}_N\Psi_N^{(0)}\label{17a}\ee
 and since $\hat H^{(0)}$ is diagonal, $\Psi^{(0)}_N$ has only one
 Fock component, $\Psi_N^{(0)}=\psi_n(P,\{\xi\}_n),$ $n=0,1,2,...,$
 and the eigenvalues  $E^{(0)}_N$ contain all possible excitation
 energies of the $n$-hybrid, with the number $n$ of gluons in the
 system fixed,
 \be
 E_N^{(0)} = E_n^{(0)} (P)
 =\sqrt{\veP^2+M^2_{n\{k|}}.\label{18a}\ee
 Here $\{ k\}$ denotes the set of quantum numbers of the excited
 $n$-hybrid.

 From (\ref{15a}) one obtains in the standard way
 corrections to the eigenvalues and eigenfunctions.

 As a first step one should
  specify the  unperturbed functions $\Psi^{(0)}_N$, introducing
  the set of quantum numbers $\{k\}$ defining the excited hybrid
  state for each $n$-hybrid Fock component $\psi_n(P\{\xi\})$; we
  shall denote therefore:
  \be
  \Psi^{(0)}_N=\psi_{n\{k\}}(P,\{\xi\}_n), ~~
  n=0,1,2,...\label{19a}\ee

  The set of functions $\psi_n\{k\}$ with all possible $n$ and
  $\{k\}$ is a complete set to be used in the  expansion of the
  exact wave-function (Fock tower) $\Psi_N$:
  \be
  \Psi_N=\sum_{m\{k\}}c^N_{m\{k\}}\psi_{m\{k\}}.\label{20a}\ee
  Using the orthonormality  condition
  \be
  \int\psi^+_{m\{k\}}\psi_{n\{p\}} d\Gamma
  =\delta_{mn}\delta_{\{k\}\{p\}}\label{21a}\ee
  where $d\Gamma$ implies integration over all internal
  coordinates and summing over all indices, one obtains from
  (\ref{14}) an equation for  $c_{m\{k\}}$ and $E_N$,
  \be c^N_{n\{p\}} (E_N-E^{(0)}_{n\{p\}}) = \sum_{m\{k\}} c^N_{m\{
  k\}}V_{n\{p\}, m\{k\}}\label{22a}
  \ee
  where we have defined
  \be
  V_{n\{p\}, m\{k\}}= \int \psi^+_{n\{p\}}\hat V \psi_{m\{k\}}
  d\Gamma.\label{23a}\ee

Consider now the Fock tower built on the valence component
$\psi_{\nu\{\kappa\}}$, where $\nu $ can be any integer. For
$\nu\{\kappa\}= 0\{0\}$ this valence component corresponds to the
unperturbed hadron with minimal number of valence particles. For
higher values of $\nu\{\kappa\}$ the Fock component
$\psi_{\nu\{\kappa\}}$  corresponds to the hybrid  with $\nu$
gluons which after taking into account the interaction is "dressed
up" and acquires all other Fock components, so that the number $N$
in (\ref{20a}) contains the "bare number" $\nu\{\kappa\}$ as its
part $N=\nu\{\kappa\},...$ (at least for small perturbation $\hat
V$).

One can impose on $\Psi_N$ the orthonormality conclusion
\be
\int\Psi^+_N\Psi_M d\Gamma = \sum_{m\{k\}} c^{N*}_{m\{k\}}
c^M_{m\{k\}} = \delta_{NM}.\label{24} \ee

Expanding now in powers of $\hat V$, one has
\be
c_{m\{k\}}^{N(\nu\{\kappa\})}= \delta_{m\nu}
\delta_{\{k\}\{\kappa\}}+ c^{N(1)}_{m\{k\}}+
c^{N(2)}_{m\{k\}}+...\label{25}\ee
\be
E_{N(\nu\{\kappa\})}=E^{(0)}_{\nu\{\kappa\}}+
 E^{(1)}_N+E^{(2)}_N+...\label{26}\ee

 It is easy to see that $E^{(1)}_N\equiv 0$, while for $c^{(1)}$
 one obtains from (\ref{22a}) the standard expression
 \be
 c^{N(1)}_{n\{p\}}=
 \frac{V_{n\{p\},\nu\{\kappa\}}}{E^{(0)}_{\nu\{\kappa\}}-E^{(0)}_{n\{p\}}}.\label{27}\ee
 In what follows we shall be  interested in  the high Fock
 components, $\nu+l,\{k\}$, obtained by adding $l$ gluons to the
 valence component $\nu\{\kappa\}$. Using (\ref{22a}) and
 (\ref{25}) one obtains
$$
 c^{N(
\nu\{\kappa\})}_{\nu+l,\{k\}}= \sum_{\{k_1\}...\{k_l\}}
\frac{V_{\nu+l\{k\},\nu+l-1\{k_1\}}}{E^{(0)}_{\nu\{\kappa\}} -
E^{(0)}_{\nu+l\{k\}}}\frac{V_{\nu+l-1\{k_1\},\nu+l-2\{k_2\}}}{E^{(0)}_{\nu\{\kappa\}}
- E^{(0)}_{\nu+l-1\{k_1\}}}...$$ \be \frac
{V_{\nu+1\{k_l\},\nu\{\kappa\}}}{E^{(0)}_{\nu\{\kappa\}} -
E^{(0)}_{\nu+1\{k_l\}}}+O(V^{l+2}).\label{28}\ee

Since $\hat V$ is proportional to $g$, one obtains in (\ref{25})
the perturbation series in powers of $\alpha_s$ for $c^N$ and
hence for $\Psi_N$ (\ref{20a}). One should note that
$\alpha_s(Q^2)$ is the background coupling constant, having the
property of saturation for positive $Q^2$ \cite{3} and the
background perturbation series has no Landau ghost pole and is
defined in all Euclidean region of $Q^2$.

The estimate of the mixing between meson and hybrid was done
earlier in the framework of the  potential model for the meson in
\cite{9}. In \cite{21} the mixing between hybrid, meson and
glueball states was calculated in the framework of the present
formalism and we shortly summarize the results. One must estimate
the matrix element (\ref{23a}) between meson and hybrid wave
functions taking the operator $\hat V$ in the form of (\ref{10a}),
where the operator of gluon emission at the point $(\vex, 0)$ can
be approximated as
 $$
 a_\mu(\vex, t)
=\sum_{\vek,\lambda}\frac{1}{\sqrt{2\mu(\vek)V}}\times $$
\be
\times [\exp({i\vek\cdot \vex-i\mu t)}e^{(\lambda)}_\mu c_\lambda
(\vek)+e^{(\lambda)}_\mu c^+_\lambda(\vek) \exp({-i\vek\cdot
\vex+i\mu t})]
 \label{42}
 \ee

 Omitting for simplicity all polarization vectors and
 spin-coupling coefficients which are  of the order of unity, one
 has the matrix element
 \be
 V_{Mh} = \frac{g}{\sqrt{2\mu_g}} \int \varphi_M (\ver
 )^\mu\psi_h^+ (0, \ver) d^3r \label{43}\ee
 where $\varphi_M(\ver), ~~^\mu\psi_h(\ver_1,\ver_2)$ are meson
 and hybrid wave functions respectively, and in (\ref{43}) it is
 taken into account that the gluon is emitted (absorbed) from the
 quark position.

 Using realistic Gaussian approximation for the wave functions in
 (\ref{43}) one obtains the estimate \cite{21}
 \be
 V_{Mh}\approx g \cdot 0.08~{\rm GeV}.\label{44}\ee
 A similar estimate is obtained in \cite{20} for the
 hybrid-glueball mixing matrix element, while the meson-glueball
 mixing is second-order in (\ref{44}).

 Hence the hybrid admixture coefficient (\ref{27}) for the meson
 is
 \be
 C_{Mh} =\frac{V_{Mh}}{E^{(0)}_M-E^{(0)}_h} =\frac{V_{Mh}}{\Delta
 M_{Mh}}\label{45}\ee
 and for the ground state low-lying mesons when $\Delta M_{Mh
 }\sim 1$ GeV it is small, $C_{Mh}\sim 0.1-0.15$, yielding a 1-2\%
 probability. However for higher states in the region $M_M\ga 1.5$
 GeV, the mass difference $\Delta M_{Mh}$ of mesons and hybrids
 with the same quantum numbers can be around 200 MeV,  and the
 mixing becomes extremely important, also for meson-glueball
 mixing, which can be written as
 \be
 C_{MG} =\sum_h \frac{V_{Mh} V_{hG}}{\Delta M_{Mh} \Delta
 M_{hG}}\label{46}\ee
 and $V_{Mh} \sim V_{hG}$. It is clear that the iterative  scheme
 described above can be useful only because hybrid excitation by
 one additional gluon "costs" around 1 GeV  increase in mass,
 hence the coefficient  $c_n^N$ (\ref{27}) can be small.

\section{Hybrid states and DIS}

 As was stressed in the previous section, in the large $N_c$ limit
 the higher Fock components which are excited by the external
 current (or incident hadron) are the multihybrid (or $n$-hybrid), states. It is
 convenient to consider these states in the light-cone formalism,
 following the line of derivation and most notations in \cite{37}.

Consider the $n$-hybrid with quark at the point $z^{(a)}_\mu$,
antiquark at the point $z^{(b)}_\mu$ and gluons at the points
$z^{(k)}_{\mu, k=1,...n}$. We also define
$\rho^{(i)}=z^{(i)}-z^{(i-1)}$, with $z^{(0)}\equiv z^{(a)}$ and
$z^{(n+1)}\equiv z^{(b)}$.

The action is
\be
A=K+\sigma S_{\min}\label{A3.25}\ee where the kinetic operator $K$
and the minimal area $S_{\min}$ can be written as $$
K=\frac{m^2_a}{2\mu_a} +\frac{m^2_b}{2\mu_b} +\frac12 \int^T_0
dz_+ [\mu_a ((\dot z_1^{(a)})^2+2\dot z^{(a)}_-) +\mu_b ((\dot
z_1^{(b)})^2+2\dot z^{(b)}_-)$$
\be
+\sum^n_{i=1}\mu_i((\dot z_1^{(i)})^2+2\dot
z^{(i)}_-)]\label{A3.26}\ee
\be
\sigma S_{\min}=\frac12\int^T_0 dz_+ \sum^{n+1}_{i=1} \int^1_0
d\beta_i\left [\nu_i \left((\dot w^{(i)})^2-\frac{(\dot
w^{(i)}w'^{(i)})^2}{(w'^{(i)})^2}\right)+\frac{\sigma^2(w'^{(i)})^2}{\nu_i}\right]\label{A3.27}\ee
 and we have defined
\be
w^{(i)}_\mu =z_\mu^{(i-1)} (1-\beta_i) +\beta_iz_\mu^{(i)}, ~~
\dot w^{(i)}=\dot z^{i-1)}(1-\beta_i) + \beta_i\dot z^{(i)},
\label{A3.28}\ee
\be
w'^{(i)}=z^{(i)} - z^{(i-1)}\equiv \rho^{(i)}.\label{A3.29}\ee

As in \cite{37} we introduce the total momentum $P_+$,
\be
P_+ = \sum^n_{i=1} \mu_i+\sum^{n+1}_{i=1}\int^1_0 \nu_i d\beta
 + \mu_a+\mu_b.
\label{A3.30}\ee  At this point one can make a new important step
and introduce  the parton's quota $x_i$ of the total momentum
$P_+$ to be associated with the Feynman variables $x_i,~~ i=1,...,
n$, $x_a$ and $x_b$, which  can be written as
\be
x_i=\frac{\mu_i+\int^1_0\nu_i\beta d \beta +\int^1_0 \nu_{i+1}
(1-\beta) d\beta}{P_+}\label{A3.31}\ee
\be
x_a\equiv x_0 = (\mu_a+ \int^1_0\nu_1 (1-\beta) d\beta)
/P_+\label{A3.32}\ee

\be
x_b\equiv x_{n+1} = (\mu_b+ \int^1_0\nu_{n+1} \beta d\beta) /P_+.
\label{A3.33}\ee

One can notice that $x_i$ consists of three pieces: 1) the $(+)$
-component of momentum of the valence gluon $(\mu_i)$, 2) the
$(+)$ momentum of the preceeding piece $(i)$ of string weighted
with the factor $\beta$ which takes into account that the string
$(i)$ is deformed by the motion of the gluon $(i)$ while another
end of the string is fixed 3)  the $(+)$ momentum of the string
$(i+1)$ weighted with the factor $(1-\beta)$ taking into account
motion of the $(i\neq 1)$ string due to the $i$-th gluon. Note
that the parameter $\beta$ in all strings $(i)$, $i=1,... n+1$
grows in one direction, e.g. from the  left to the right.

In this way the momentum of each piece of the string $(i)$ is
shared by two adjacent gluons: $(i-1)$ and $(i)$, so that each
parton quota $x_i$ contains momentum of the parton (gluon or quark
or antiquark) itself and of pieces of adjacent strings.

These results imply  several nontrivial consequences. First of
all, one can see that the einbein factors $\mu_i$, which played in
the c.m. system the role of constituent mass (energy) of gluon and
quarks, and $\nu_i$ played the role of energy density along the
string, in the l.c. system they enter directly the Feynman
parameters of gluons. In this way one can for the first time  see
the connection of the standard constituent quark (gluon) picture
with the parton picture and calculate as in (\ref{A3.32}),
(\ref{A3.33}) parton parameters through the (Lorentz boosted)
constituent energies  of quarks and gluons. Secondly in the  l.c.
wave-function of the n-hybrid the average values of $\mu_i$ and
$\nu_i$ are equal for large $n, \bar \mu_i=\bar \nu_i,
i=1,2,...n$, while for $n=1$ one obtains $\bar\mu_g = \sqrt{2}
\bar \mu_q = \sqrt{2} \bar \mu_{\bar q}$ \cite{38}.

Hence gluons carry  more momentum on average than quarks in the
$n$-hybrid state. Therefore one expects that in DIS at large
enough energy when the $n$-hybrid component of the hadron
wave-function is excited, the contribution of gluons to the
momentum sum rule and to the proton spin should be dominant. This
expectation is consistent with the experimental data. Thus in the
neutrino-isoscalar scattering the momentum sum rules for the quark
part of $F_2(x, Q^2)$ yield \cite{39} 0.44$\pm$ 0.003, which
implies that gluons carry more than 50\% of the total momentum.

For the  proton spin one has the relation \cite{40}
\be
\frac12=\frac12 \Delta \Sigma (\mu) + L_q (\mu) + J_g
(\mu)\label{53}\ee where the quark sigma-term experimentally is
$\Delta \Sigma (\mu=1 {\rm ~GeV}) = 0.2 \pm 0.1$, and the most
part of the difference between the l.h.s. and the  r.h.s. is
presumably due to the gluon spin contribution $J_g(\mu)$.  The
detailed estimates of hybrid contributions to DIS and high-energy
scattering will be published elsewhere \cite{38}.

\section{Conclusions}

It was explained above that the Field Correlator Method is a
powerful tool for investigation of all nonperturbative effects in
QCD. In particular it provides a natural mechanism of confinement,
compatible with all lattice  data, and explains the close
connection of confinement and chiral symmetry breaking. The
spectrum of mesons, glueballs and hybrids is calculated with
$\sigma, \alpha_s$ and current masses as the only fixed parameters
used and this spectrum is  in good agreement with  lattice data
and experiment. The latest development concerns the dominant role
of hybrids in DIS and high-energy scattering and here the first
qualitative results are consistent with experimental evidence. The
author is grateful to the organizers of the Pomeranchuk
International Conference for their  excellent job,  and to
A.M.Badalian, K.G.Boreskov, A.B.Kaidalov and O.V.Kancheli for many
stimulating discussions.

The partial support of the INTAS grants 00-110 and 00-366 is
gratefully acknowledged.

\end{document}